\documentclass[conference]{IEEEtran}
\IEEEoverridecommandlockouts
\usepackage{cite}
\usepackage{amsmath,amssymb,amsfonts}
\usepackage{algorithmic}
\usepackage{graphicx}
\usepackage{textcomp}
\usepackage{xcolor}
\usepackage{subcaption}
\usepackage{float} 
\usepackage{url}
\usepackage{fancyhdr}

\def\BibTeX{{\rm B\kern-.05em{\sc i\kern-.025em b}\kern-.08em
    T\kern-.1667em\lower.7ex\hbox{E}\kern-.125emX}}

\fancypagestyle{firstpage}{
    \fancyhf{}
    \fancyhead[L]{\small \textcolor{red} {This paper has been accepted in 1st International Conference on AIML-Applications for Engineering \& Technology (IEEE ICAET-2025). Published version of the paper will be available in IEEE Xplore very soon.}}
    \fancyhead[C]{}
    \fancyhead[R]{}

}

\begin{document}

\title{Rice Grain Size Measurement using Image Processing  \\
}


\author{\IEEEauthorblockN{1\textsuperscript{st} Ankush Tyagi}
\IEEEauthorblockA{\textit{Software Development Manager} \\
\textit{Ericsson}\\
Austin, Texas, USA \\
ankush.tyagi@ericsson.com}
\and
\IEEEauthorblockN{2\textsuperscript{nd} Dhruv Motwani}
\IEEEauthorblockA{\textit{Head of GenAI} \\
\textit{Avahi}\\
San Francisco, CA, USA \\
dhruv.motwani@avahitech.com}
\and
\IEEEauthorblockN{3\textsuperscript{rd} Vipul Dabhi }
\IEEEauthorblockA{\textit{Department of Information Technology} \\
\textit{Dharmsinh Desai University}\\
Nadiad, India \\
vipuldabhi.it@ddu.ac.in
}
\and
\IEEEauthorblockN{4\textsuperscript{th} Harshadkumar Prajapati }
\IEEEauthorblockA{\textit{Department of Information Technology} \\
\textit{Dharmsinh Desai University}\\
Nadiad, India \\
prajapatihb.it@ddu.ac.in
}
}

\maketitle

\thispagestyle{firstpage}

\begin{abstract}
The rice grain quality can be determined from its size and chalkiness. The traditional approach to measure the rice grain size involves manual inspection, which is inefficient and leads to inconsistent results. To address this issue,  an image processing based approach is proposed and developed in this research. The approach takes image of rice grains as input and outputs the number of rice grains and size of each rice grain. The different steps, such as extraction of region of interest, segmentation of rice grains, and sub-contours removal, involved in the proposed approach are discussed. The approach was tested on rice grain images captured from different height using mobile phone camera.
The obtained results show that the proposed approach successfully detected 95\% of the rice grains and achieved 90\% accuracy for length and width measurement.   
\end{abstract}


\section{Introduction}
In today's market, consumers are increasingly conscious about the quality of the rice grains they purchase, especially with growing reports of adulteration. Rice shape (dimension) and chalkiness are important parameters to determine its quality and price. Manual analysis of rice grain samples is not only time-consuming and complex but also prone to errors due to the subjective nature of human perception. 

Generally, an inspector performs the visual inspection of rice grain to determine its dimension (length and width). The inspector uses a ruler or caliper to measure the dimension of rice grain. The accurate measurement of dimension of rice grains is useful to determine its quality and pricing. Conventional methods of measuring the length and width of rice grains is time consuming and prone to error. Moreover, the dimension measurement, of same rice grains, from different inspectors may vary and affected by work pressure, eye sight issue, weather and lighting conditions. Therefore, an automatic, rapid and accurate measurement method to measure rice grain dimensions is need of a day. Researchers have applied different image processing techniques to measure the dimension of rice grains. These researchers tried to improve the effectiveness and accuracy of rice grains measurement. However, these methods produce different results for different environmental conditions such as background image and light conditions. We developed image processing based technique to measure the dimension of rice grains. We also discussed different problems and their solutions to improve the accuracy of rice measurement. Moreover, our approach accurately measure and produce same result for rice grain images captured from different heights.

The rest of the paper is organized as follows: Section 2 presents related work. Next section discusses the flowchart and image processing steps involved in rice grain size measurement. In Section 4, we briefly discusses the design challenges faced and the experimental results obtained using the proposed approach. Finally, Section 5 concludes the paper.

\section {Related Work}
Significant research has focused on grading various food grains, with some studies combining image processing and machine learning to classify rice grains \cite{cinar2022identification, kurade2023automated}. These methods use image processing to extract features like color, texture, and shape, which are then processed by machine learning algorithms. Other approaches grade rice grains based on width, length, length-width ratio, and area but rely on images captured from a fixed height. These methods fail when images are taken from varying heights. This research aims to develop an image processing approach to accurately measure rice grain size from images captured at different heights.

The colour of seed holder (canvas color) can have a considerable effect on the rice grain size measurement. In \cite{ruslan2018effect}, authors performed experiments with four (Black, Blue, Green and Red) different colours of rice grain holder. They reported that black and blue colour rice grain holder had lowest percentage of error in measurement of rice length and width. Backlight photography was used in \cite{feng2023size} to capture a grayscale image of a group of rice grains. The authors \cite{feng2023size} reported the performance of the proposed method on three key metrics, the total number of rice grains; grain length measurement; and grain width measurement, using Coefficient of Determination (R2) and Root Mean Square Error (RMSE).

Authors \cite{cinar2022identification} used a dataset of 75,000 rice grains images, of 5 rice varieties, classification of rice grains. The pre-processed images were used for extraction of 106 features (12 morphological, 4 shape and 90 color features). The extracted features were given as input to different classification algorithms (K-NN, DT, LR, MLP, RF and SVM). The authors \cite{cinar2022identification} reported that MLP outperforms, with 99.91\% accuracy, as compared to other algorithms. A Raspberry-Pi based image processing module was designed in \cite{kurade2023automated} to detect adulteration of low-quality rice grains with high-quality rice grains. The color, texture, and geometry features of rice grains were extracted and given as inputs to different machine learning models. It was found that Random Forest (RF) outperforms other ML models with accuracy of 76\%.

Authors \cite{lin2017determination} reviewed various image-based methods for rice classification and grading. They categorized these methods into five approaches: geometric, statistical, supervised, unsupervised, and deep learning. They found deep learning techniques to be the most promising. However, they highlighted the challenge of building accurate deep learning models, noting the difficulty in developing large rice grain datasets in uncontrolled environments. The data sets with non-uniform lighting, occlusion among rices and indistinguishable rice grains will continue to pose challenges for future research. An image processing based approach for counting the number of rice grains and grading these grains based on length, width and length-width ratio was proposed in \cite{singathala2023quality, kuchekar2018rice, karunasena2020machine}. However, to locate the boundary of rice grains, \cite{kuchekar2018rice} used canny whereas \cite{singathala2023quality} applied sobel edge detection algorithms. The authors \cite{karunasena2020machine} reported an accuracy of 85\% on measurement of length of rice grains and more than 90\% for L/W (length-width) ratio.  


\section{Proposed Method}
\begin{figure}[ht]
    \centering
    {\includegraphics[width=1.0\linewidth]{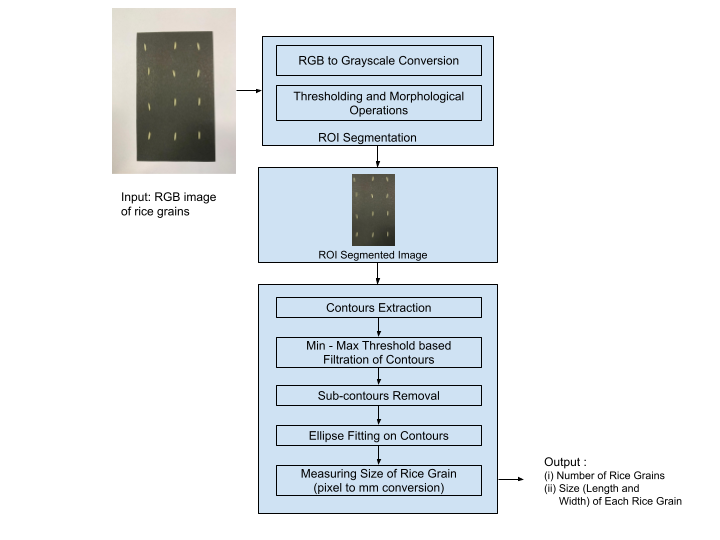}}
    \caption{Flowchart of the proposed approach for rice grain size measurement}
    \label{Fig:Flowchart}
\end{figure}

This section presents the different image processing steps to detect the number of rice grains, followed by measuring the length and width of detected rice grains. These steps are presented in Figure \ref{Fig:Flowchart}. The proposed approach consists of three essential steps: (i) segmentation of Region of Interest (RoI), (ii) extraction and filtration of contours that represent rice grains and (iii) measurement of contours with reference to the known dimension of the RoI. While the primary focus of this work is on rice grains measurement, the steps of the proposed approach can be adapted for use in other agricultural sectors where similar quality assessments are needed. The approach could be extended to analyze other grains, such as wheat, barley, and millet, by calibrating the system to recognize unique shape, size, and structural characteristics. The details of the proposed approach are discussed in the following subsections. 

\subsection{Image Acquisition}
The image acquisition is the most important step in detection and size measurement of the rice grains as it directly affects the final result. The rice grains were placed uniformly without touching each other on a black color paper, referred as canvas. The type of rice used in this research is Basmati. The images of rice grains were taken using different mobile phone cameras in RGB format. Total 30 images were acquired using mobile phone camera. The dimension of captured image was 4032 X 3024 pixels in JPG format. However, images in PNG and JPEG format can also be processed using the proposed approach. To evaluate the proposed approach in different light conditions, images were taken (i) with mobile flash light off and (ii) with mobile flash light on. Moreover, these images were taken from different heights. The objective behind taking the images from different height was to evaluate the performance of the proposed approach against variation in height. 

Accurate measurement of rice grain size from images captured at different height is a challenging problem. To overcome the issue of height, use of reference object in captured image was necessary. We have used black color paper canvas as a reference object. The reference object selected for this purpose is a black rectangular canvas with know dimensions. Using this reference object, the pixels-per-millimeter ratio is calculated for each image. This ratio is used as a conversion factor, to convert the measured size of rice grains in pixels to mm. The use of reference object ensures that the measurements are reliable and comparable across different images.

\begin{figure}[ht]
    \centering
    \begin{minipage}{0.4\linewidth}
        \centering
        \includegraphics[width=\linewidth]{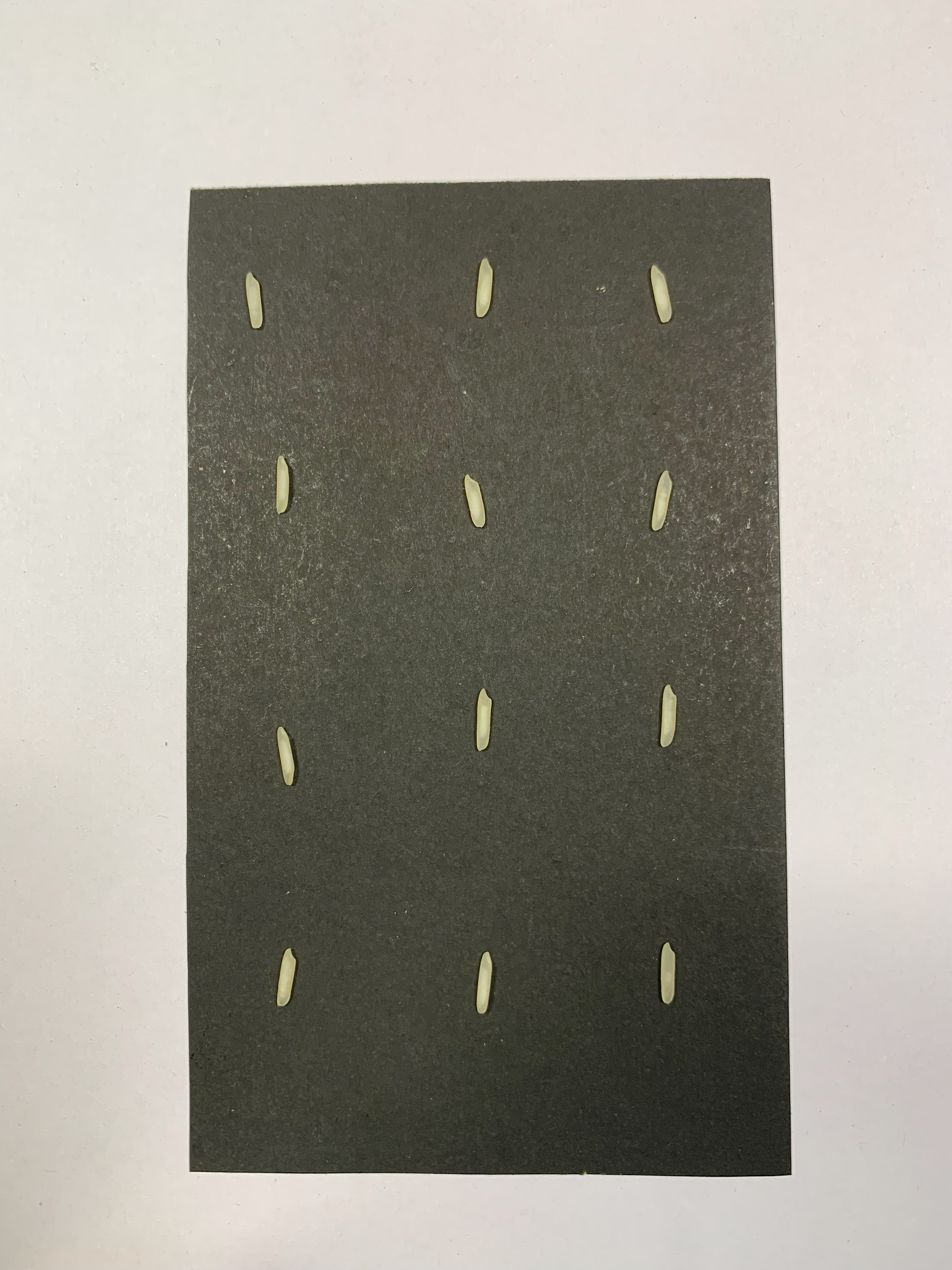}
        \subcaption{Flash Light Off}
    \end{minipage}
    \hspace{0.02\linewidth}
    \begin{minipage}{0.4\linewidth}
        \centering
        \includegraphics[width=\linewidth]{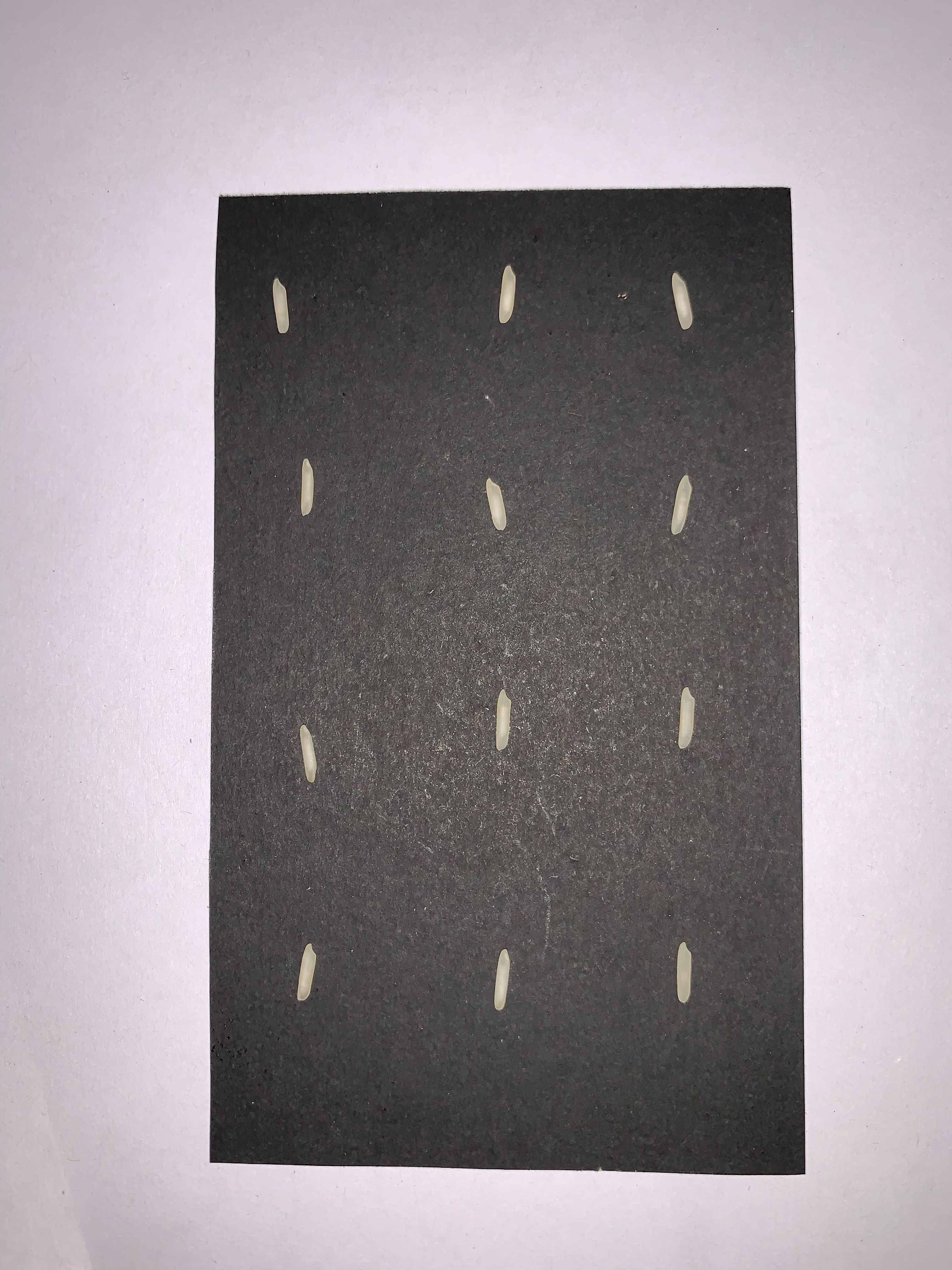}
        \subcaption{Flash Light On}
    \end{minipage}
\caption{Images taken using Mobile Camera}
\end{figure}

\subsection{Processing of Rice Grain Image}
The original RGB (red, green, blue) image was converted to grayscale image.  A Gaussian blur filter kernel of size 5X5 was applied to the grayscale image to reduce noise and smooth the image. 

\subsubsection{Canvas Detection}
To differentiate the canvas from the background, Otsu's thresholding was applied. The Otsu's thresholding gave the optimal value for threshold to segment the foreground and background and to generate the binary image.

Following thresholding, the contours from the image were detected. To ensure that only canvas was selected, only those contours which were either square or rectangular in shape were selected. However, during the canvas extraction process, a common issue encountered was the appearance of white pixels along the border of the canvas. These white pixels may affect the accuracy of rice grain measurement and need to be removed. To remove these white pixels, the extracted canvas region was cropped by 5\% from the border, effectively removing the unwanted white pixels and refining the region of interest. This step ensured accurate extraction of canvas region.


\subsubsection{Outliers in Rice Grain Size Measurement}
The adaptive thresholding was applied on extracted canvas region (region of interest). The adaptive thresholding was selected to tackle uneven lighting conditions and to distinguish rice grains from the black background. After that contour detection and filteration was performed on the binary image using OpenCV library.    

\begin{figure}[ht]
    \centering
    \begin{minipage}{0.4\linewidth}
        \centering
        \includegraphics[width=\linewidth]{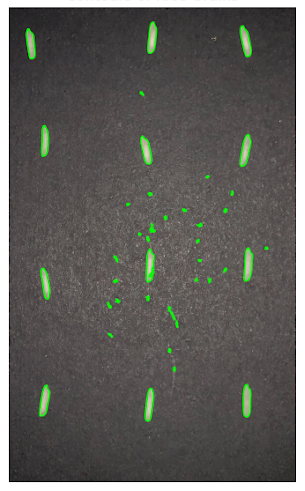}
        \subcaption{Flash Light Off}
    \end{minipage}
    \hspace{0.02\linewidth}
    \begin{minipage}{0.4\linewidth}
        \centering
        \includegraphics[width=\linewidth]{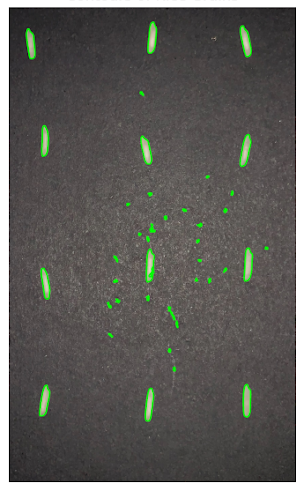}
        \subcaption{Flash Light On}
    \end{minipage}
\caption{Outliers in Rice Grain Size Measurement}
\end{figure}

\subsection{Selection of Contour Filtration Approach}
During the process of measuring rice grains dimension using image processing, one of the issues encountered is the detection of small contours which do not correspond to actual rice grains. Due to these erroneously detected contours, the proposed method was giving more number of rice grains as compared to actual number of rice grains in the image. Therefore, to remove smaller contours which are likely to be noise and filter the relevant contours, the following approaches were applied. 

\subsubsection{Using Median Value based Thresholding}
To handle the issue of smaller contours, the median value (of length and width) based thresholding was applied. The median value of (length and width) all detected contours was calculated. The upper and lower threshold values were used to filter out contours.

The median value based thresholding approach performed well in cases where the number of correctly detected rice grain contours was predominant but fails in cases where the number of erroneously detected contours was more as compared to correctly detected rice grain contours. In such cases, the median value tended to shift towards the dimensions of these erroneous contours. Consequently, the approach, instead of eliminating the incorrect contours, ended up preserving them and possibly discarding the true rice grain contours.

\subsubsection{Using Min Max Value based Thresholding}
To overcome the issue of wrongly detected contours, a thresholding approach based on predefined minimum and maximum values for length and width of rice grain was applied. The minimum and maximum length of 4 mm and 15 mm and the minimum and maximum width of 1 mm and 4 mm was considered for rice grain. This approach produced consistent result by ensuring that only contours corresponding to actual rice grains were retained.

\section{Results and Discussion}
The ground-truth size (length and width) of rice grains were measured manually using ruler. The measured dimensions were compared with the dimensions obtained using the proposed method. We have considered twelve rice grains. The mean absolute error between the manually measured dimensions and obtained dimensions were calculated for each rice grain. 




    \begin{figure}[ht]
        \centering
        \begin{minipage}{0.4\linewidth}
            \centering
            \includegraphics[width=\linewidth]{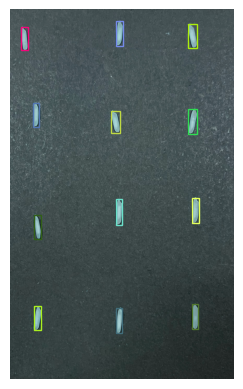}
            \subcaption{Flash Light Off}
        \end{minipage}
        \hspace{0.01\linewidth}
        \begin{minipage}{0.4\linewidth}
            \centering
            \includegraphics[width=\linewidth]{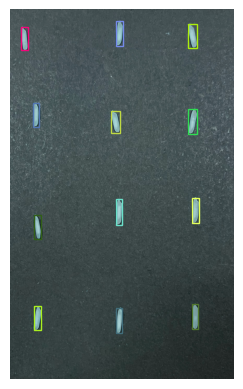}
            \subcaption{Flash Light On}
        \end{minipage}
    \caption{Rectangular Bounding Box Enclosing Rice Grains}
    \end{figure}

\subsection{Problem of Sub-regions in Rice Grain Measurement}

\begin{figure}[ht]
    \centering
    {\includegraphics[width=0.5\linewidth]{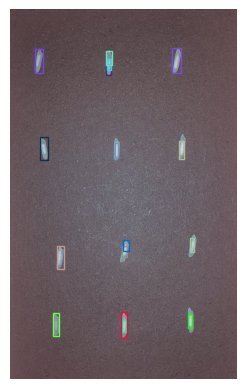}}
    \caption{Problem of Sub-regions}
    \label{fig:sub-regions}
\end{figure}

During the experiments, a significant challenge encountered was related to
generation of multiple sub-contours or sub-regions for a single rice grain. For example, in the Figure \ref{fig:sub-regions}, for rice grains number 2 and 11, the approach detected multiple sub-contours. These sub-contours represent the smaller region of the rice grain and 
were part of the primary contour. The image processing approach detected these sub-contours separately and misinterpreted them as individual rice grain. Due to which, the approach was predicting more number rice grains as compared to actual number of rice grains in the image. Moreover, it was observed that the problem of sub-contours occurred in images where the contrast between the rice grain and background was not well-defined.

To overcome this challenge, the sub-contour removal algorithm is designed. The algorithm checks whether one contour is fully enclosed within another contour or not. The algorithm uses coordinates, height and width of detected contours to determine the sub-contours. The algorithm helps in improving the accuracy of rice grain dimensions measurement by reducing the false detected rice grains.

\subsubsection{Rice Grain Enclosing Ellipse}
To approximate the length and width of a rice grain, the rectangular region was created around the detected contour. The dimension of the rectangular region gives the approximate length and width of rice grain. This method worked well for rice grains that were aligned either vertically or horizontally. However, when the rice grains were positioned at an angle, the rectangular bounding box failed to accurately represent the grain's true dimensions. Specifically, the width of the rectangle often exceeded the actual width of the rice grain, leading to overestimation. This problem occurred because the bounding box is axis-aligned and does not consider the orientation or angle at which rice grain was placed.

An ellipse is constructed around the detected contour of each rice grain to accurately determine its size. The best-fit ellipse is used to estimate the length and width of the rice grain, where the length corresponds to the major axis of the ellipse, and the width corresponds to the minor axis. This approach precisely approximated rice grain size by considering the orientation and shape of grain.

\subsection{Rice Grain Size Measurement}
The pixel values along the length and width of the rectangle were measured. These values then converted to millimeters using pixel to mm ratio, which was determined earlier.

\subsection{Result}
\begin{figure}[ht]
    \centering
    {\includegraphics[width=0.5\linewidth]{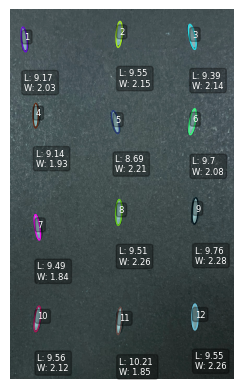}}
    \caption{Detected Rice Grains with their size, obtained using the Proposed Approach}
    \label{Figure:results}
\end{figure}

Figure \ref{Figure:results} presents the number of rice grains detected and the size of each rice grain obtained with the proposed approach. It was observed that the approach was able to detect all rice grains successfully. Table \ref{Table:results} compares the manually measured rice grain sizes with those obtained using the proposed approach on a sample image. After testing on 30 images, the approach successfully detected 95\% of the rice grains and provided 90\% accuracy for length and width measurements.

\begin{table}[!ht]
\begin{center}
\caption{The actual and predicted (with our proposed approach) length and width of rice grains (mm) }
\label{Table:results}
\setlength{\tabcolsep}{8pt} 
\renewcommand{\arraystretch}{1} 
\begin{tabular}{|c c c c c|} 
\hline
\textbf{Sample} & \multicolumn{2}{ c }{\textbf{Proposed Approach}} & \multicolumn{2}{ c |}{\textbf{Ground truth (Ruler)}}  \\ 
\cline{2-5}
& \textbf{Length} & \textbf{Width} & \textbf{Length} & \textbf{Width} \\
\hline

1 & 9.17 & 2.03 & 8.79 & 2.02  \\  
2 & 9.55 & 2.15 & 8.67 & 1.91  \\  
3 & 9.39 & 2.14 & 8.92 & 2.04  \\  
4 & 9.14 & 1.93 & 8.54 & 2.08  \\  
5 & 8.69 & 2.21 & 8.32 & 2.12  \\  
6 & 9.70 & 2.08 & 8.97 & 1.94  \\  
7 & 9.49 & 1.84 & 8.51 & 2.07  \\  
8 & 9.51 & 2.26 & 8.56 & 2.00  \\   
9 & 9.76 & 2.28 & 8.50 & 1.98  \\   
10 & 9.56 & 2.12 & 8.30 & 1.91  \\  
11 & 10.21 & 1.85 & 9.42 & 2.01  \\  
12 & 9.55 & 2.26 & 9.19 & 1.88  \\  \hline

\end{tabular}
\end{center}
\end{table}




\section{Conclusion and Future Work}
An automatic rice grain size measurement approach based on image processing  was proposed and implemented. The following challenges, encountered during the development were discussed in detail: (i) selection of contour filtration approach and (ii) selection of contour measurement approach. The proposed approach was tested on rice grain images. The performance of the approach was measured based on (a) number of rice grains detected vs actual number of rice grains (b) difference between the predicted dimensions (length and width) and actual dimensions of rice grains. The proposed approach achieved 95\% accuracy for rice grains detection and 90\% accuracy for rice grain length and width measurement. 
In future, the proposed approach will be extended to measure the dimensions of rice grains in the following two scenario : (i) rice grains placed at the border of the canvas (ii) rice grains touching with each other. 


\bibliographystyle{ieeetr}
\bibliography{refs} 

\vspace{12pt}

\end{document}